\documentclass[12pt]{article}

\ifx\pdfoutput\undefined
\usepackage[dvips,bookmarks]{hyperref}  
\else
\usepackage{hyperref}   
\fi

\hypersetup{colorlinks,bookmarksopen,bookmarksnumbered,citecolor=blue,
        pdfstartview=FitH}
\def\hhref#1{\href{http://arxiv.org/abs/hep-th/#1}{hep-th/#1}} 
\def\mhref#1{\href{mailto:#1}{#1}}              

\def\bop#1{\setbox0=\hbox{$#1M$}\mkern1.5mu
    \vbox{\hrule height0pt depth.04\ht0
    \hbox{\vrule width.04\ht0 height.9\ht0 \kern.9\ht0
    \vrule width.04\ht0}\hrule height.04\ht0}\mkern1.5mu}
\def\bo{{\mathpalette\bop{}}}                        
\def\frac#1#2{{\textstyle{#1\over#2}}}     

\parskip=\medskipamount         
\thicklines             
\begin{document}

\begin{titlepage}
\begin{flushright}
YITP-SB-03-53\\ October 7, 2003\\
\end{flushright}

\begin{center}
{\centering {\Large\bf Gauge-covariant vertex operators \par} }

\vskip 1cm {\normalsize\bf Haidong Feng\footnote{E-mail address:
\mhref{hfeng@ic.sunysb.edu}}, and Warren Siegel\footnote{E-mail
address: \mhref{siegel@insti.physics.sunysb.edu}}} \\ \vskip 0.5cm

{\it C.N. Yang Institute for Theoretical Physics,\\
 State University of New York, Stony Brook, 11790-3840 \\}

\end{center}

\begin{abstract}
We derive Yang-Mills vertex operators for (super)string theory whose BRST invariance requires only the free gauge-covariant field equation and no gauge condition.  Standard conformal field theory methods yield the three-point vertices directly in gauge-invariant form.
\end{abstract}

\end{titlepage}

\section{Introduction}

In string theory the usual vertex operators are not gauge invariant.
For example, the condition $k\cdot \epsilon=0$ is imposed on the open-string vertex operator
$\epsilon \cdot \partial X e^{ik \cdot X}$
for the requirement of conformal weight
1 (see, e.g., \cite{string theory}).  On the other hand, nonlinear sigma models
directly give gauge invariant results,
but only order-by-order in $\alpha'$, and thus not the complete scattering
amplitude \cite{Fradkin:1985}.

In string theory there are two kinds of vertex
operators in
any amplitude -- an integrated one $\oint W$ and an unintegrated one $V$:
\begin{equation}
{\cal A} = \langle V V V \textstyle{\oint}\thinspace W \cdots \textstyle{\oint}\thinspace W\rangle
\end{equation}
where for a gauge vector
\begin{equation}
W = A(X) \cdot \partial X
\end{equation}
or in Fourier expansion
\begin{equation}
W = \epsilon \cdot \partial X e^{ik \cdot X}
\end{equation}
Then, using the BRST operator ``$Q$", we can find the unintegrated
operator $V$ \cite{bigpicture}:
\begin{equation}
[Q,\textstyle{\oint}\thinspace W\} = 0 \quad\Rightarrow\quad
[Q, W \} = \partial V \quad\Rightarrow\quad
[Q, V \} =0
\end{equation}

In this paper, we will use the BRST operator and the integrated
vertex operator of the massless vector to construct a
gauge-covariant unintegrated vertex operator. This allows
computation of amplitudes without fixing the gauge on external
lines. Then the amplitude between 3 gauge bosons is computed and
found to be the same as the 3-point YM vertex found from a
gauge-invariant action. Also, the conformal symmetry of the
amplitude is studied to show that the result follows from any
3-string vertex for string field theory \cite{E. peskin:1988}.
This formalism is then generalized to the Neveu-Schwarz case.

\section{Bosonic vertex}

For the bosonic string, we calculate
\begin{equation}
Q = \oint \frac{1}{2\pi i} dz (-\frac{1}{4 \alpha'} c
\partial X \cdot \partial X + b c \partial c )
\end{equation}
\begin{equation}
[Q, \epsilon \cdot \partial X e^{ik \cdot X} ] = \oint
\frac{1}{2\pi i} dz' (-\frac{1}{4 \alpha'} c
\partial' X \cdot \partial' X) ~ \epsilon \cdot \partial X e^{ik \cdot X}
\end{equation}
In the upper-half complex plane, the $X$
propagator is $-2\alpha' ln
|z'-z| \eta^{\mu \nu}$, so
\begin{eqnarray}
[Q, \epsilon \cdot \partial X e^{ik \cdot X} ] = \partial (c
\epsilon \cdot \partial X ) e^{ik \cdot X} + c ( \epsilon \cdot
\partial X) (i k \cdot \partial X ) e^{ik \cdot X} \nonumber \\ + \alpha'
[( \epsilon \cdot \partial X) k^2 \partial c  e^{ik \cdot X} - (i
k \cdot \epsilon) \partial^2c e^{ik \cdot X}]
\end{eqnarray}
The first two terms come from the two ways to contract a single pair
of $X^{\mu}(z')$ and $X^{\nu}(z)$, and the last two terms from
the two ways to contract two pairs. To write it as a total
derivative, using the {\it gauge-invariant} equation of motion of the free vector
\begin{equation}\label{motion equ}
\partial^\mu F_{\mu\nu} = 0 \quad \quad
or \quad \quad k^2 \epsilon^{\mu} - k^{\mu} (k \cdot \epsilon) = 0
\end{equation}
we have
\begin{equation}
[Q, \epsilon \cdot \partial X e^{ik \cdot X} ] = \partial [c
\epsilon \cdot \partial X e^{ik \cdot X} - i \alpha' (\partial c)
(\epsilon \cdot k) e^{ik \cdot X} ]
\end{equation}

Thus we get a BRST-invariant vertex operator for the gauge vector
without gauge fixing:
\begin{equation}
V = c \epsilon \cdot \partial X e^{ik \cdot X} - i \alpha'
(\partial c) (\epsilon \cdot k) e^{ik \cdot X}
\end{equation}
\begin{equation}
or \quad V = c A \cdot \partial X - \alpha' (\partial c) (\partial\cdot A )
\end{equation}
$\partial c$ is also the vertex operator for the Nakanishi-Lautrup field $B$ \cite{NL}:
In the gauge $b_0=0$, $B=\partial\cdot A$, and the two $\partial c$ terms cancel.

Under gauge transformations,
\begin{equation}
\delta A^{\mu} = \partial^{\mu} \lambda \quad\Rightarrow\quad
\delta V = c (\partial^{\mu} \lambda) (\partial X_{\mu}) - \alpha'
(\partial c) \partial_{\mu} (\partial^{\mu} \lambda)
\end{equation}
which can be written as the commutator between $Q$ and $\lambda$
\begin{eqnarray}
[Q, \lambda] & = & \oint \frac{1}{2\pi i} dz' [-\frac{1}{4
\alpha'} c \partial' X_ (z') \cdot \partial' X (z')]
\lambda[X(z)] \nonumber \\
& = & c (\partial^{\mu} \lambda)
(\partial X_{\mu}) - \alpha' (\partial c) \partial_{\mu}
(\partial^{\mu} \lambda)  \nonumber \\
& = & \delta V
\end{eqnarray}
It is easy to see the integrated operator is gauge invariant
\begin{equation}
\delta \oint W = \oint \partial X^{\mu} \partial_{\mu} \lambda =
\oint \partial \lambda = 0
\end{equation}
so matrix elements are also:
\begin{equation}
\delta_{\lambda_1} {\cal A}_n = \langle \delta V_1 V_2 V_3
\textstyle{\oint} W \cdots \textstyle{\oint} W \rangle = \langle
[Q, \lambda_1 ] V_2 V_3 \textstyle{\oint} W \cdots
\textstyle{\oint} W \rangle = 0
\end{equation}
where the vacuum is BRST invariant, and similarly for the gauge
transformations of $V_2$ and $V_3$.

\section{Bosonic three-point amplitudes}

The simplest case is the 3-point amplitude:
\begin{eqnarray}
{\cal A}_3 & = & - \frac{i g_{YM}}{2\alpha'} \langle V_1
V_2 V_3 \rangle \nonumber
\\ & = & - \frac{i g_{YM}}{2\alpha'} \langle \prod^3_{i=1} [
\underbrace{c(y_i) \epsilon_i \cdot \partial X(y_i) e^{ik_i \cdot
X(y_i)} }_{G(y_i)} - \underbrace{i \alpha'
\partial c(y_i) (\epsilon_i \cdot k_i) e^{ik_i \cdot X(y_i)} }_{H(y_i)}] \rangle
\end{eqnarray}
Considering the lowest order of $\alpha'$, only 2 terms
contribute:
\begin{equation}
\langle \prod^3_{i=1} G(y_i) \rangle \quad and \quad \langle
G(y_i) \prod_{i \neq j} H(y_j)\rangle
\end{equation}
Using
\begin{eqnarray}
& \langle c(y_1) c(y_2) c(y_3) \rangle = y_{12} y_{13} y_{23}
\nonumber \\ & \langle \partial_{y_1} c(y_1) c(y_2) c(y_3) \rangle
= \partial_{y_1} ( y_{12} y_{13} y_{23}), \cdots
\end{eqnarray}
and the propagator between $X^{\mu}(z')$ and $X^{\nu}(z)$, and
contracting 2 of the $\partial X$'s with each other, the amplitude between 3
gauge bosons is:
\begin{eqnarray}
{\cal A}_3^{(1)} =& i g_{YM} (2\pi)^D \delta^D (\Sigma_i k_i) [ (\epsilon_1
\cdot \epsilon_2) (\epsilon_3 \cdot k_{12}) + (\epsilon_2 \cdot
\epsilon_3) (\epsilon_1 \cdot k_{23}) + (\epsilon_3 \cdot
\epsilon_1) (\epsilon_2 \cdot k_{31}) ] \nonumber \\ & \times
|y_{12}|^{2\alpha' k_1 \cdot k_2} |y_{13}|^{2\alpha' k_1 \cdot
k_3} |y_{23}|^{2\alpha' k_2 \cdot k_3}
\end{eqnarray}
where $k_{ij} = k_i - k_j$. To the lowest order in $\alpha'$, the last
factor in ${\cal A}_3$,
\begin{equation}
|y_{12}|^{2\alpha' k_1 \cdot k_2} |y_{13}|^{2\alpha' k_1 \cdot
k_3} |y_{23}|^{2\alpha' k_2 \cdot k_3} \rightarrow 1
\end{equation}
when $\alpha' \rightarrow 0$.
This amplitude corresponds to the cubic part of
\begin{equation}
\frac{1}{g_{YM}^{2}} \int d^{26} x [ -\frac{1}{4} Tr(F^{\mu \nu}
F_{\mu \nu}) ]
\end{equation}
in Yang-Mills theory, without gauge fixing (or use of a gauge condition).

With the same technology but tedious
calculation, the 3-point amplitude to second order in $\alpha'$
contributed to by
\begin{equation}
\langle G (y_1) G (y_2) G (y_3) \rangle, \quad
\langle G (y_1) G (y_2) H (y_3) \rangle, \cdots, \quad and \langle G
(y_1) H (y_2) H (y_3) \rangle, \cdots
\end{equation}
where $\cdots$ represents the permutaion of $y_i$'s, is:
\begin{eqnarray}
{\cal A}_3^{(2)} & = & 2 i \alpha' g_{YM} (2\pi)^{26} \delta^{26}
(\sum_i k_i) \times\nonumber \\ & &[ (\epsilon_1 \cdot k_2)
(\epsilon_2 \cdot k_3) (\epsilon_3 \cdot k_1) - (\epsilon_1 \cdot
k_3) (\epsilon_2 \cdot k_1) (\epsilon_3 \cdot k_2) ]
\end{eqnarray}
The $H^3$ contribution vanishes.

The remaining contributions to the amplitude, from expanding the factors
\begin{equation}
|y_{12}|^{2\alpha' k_1 \cdot k_2} |y_{13}|^{2\alpha' k_1 \cdot
k_3} |y_{23}|^{2\alpha' k_2 \cdot k_3}
\end{equation}
are just zero because
\begin{eqnarray}\label{ksquared}
& k_1 \cdot k_2 = \frac{1}{2} (k_3^2 - k_1^2 - k_2^2), \cdots
\quad \quad and \nonumber \\ & k_1^2 [ (\epsilon_1 \cdot
\epsilon_2) (\epsilon_3 \cdot k_{12}) + (\epsilon_2 \cdot
\epsilon_3) (\epsilon_1 \cdot k_{23}) + (\epsilon_3 \cdot
\epsilon_1) (\epsilon_2 \cdot k_{31}) ] = 0,  \cdots
\end{eqnarray}
by using only the {\it gauge-covariant} equation of motion (\ref{motion equ})
and momentum conservation $k_1+k_2+k_3=0$.

This ${\cal A}_3^{(2)}$ gives exactly the cubic term in the YM field:
\begin{equation}
\frac{-2i\alpha'}{3g_{YM}^{2}} Tr({F_{\mu}}^{\nu}
{F_{\nu}}^{\omega} {F_{\omega}}^{\mu} )
\end{equation}

Another simple 3-point amplitude we can calculate with this
gauge independent vertex operator is between 1 gauge boson and 2
tachyons. Using $V_t = c e^{ik \cdot X} $, to lowest order in
$\alpha'$,
\begin{equation}
-i g_{YM} g^2_0 \langle V(y_1) V_t (y_2) V_t (y_3) \rangle = -i
g_{YM} \epsilon_1 \cdot (k_2 - k_3) (2\pi)^{26} \delta^{26}
(\Sigma_i k_i)
\end{equation}
with $g_0 = (2\alpha')^{1/2} g_{YM}$ is the coupling constant for
tachyons. This corresponds to
\begin{equation}
\frac{1}{g_{YM}^{2}}[ -\frac{1}{2} Tr(D_{\mu} \phi D^{\mu} \phi) ]
\quad with \quad D_{\mu} \phi = \partial \phi - i [A_{\mu} , \phi]
\end{equation}
in quantum field theory.

One thing we can observe is that the amplitude is independent of
the conformal map to the complex plane. This can be verified by checking the
conformal transformation of the vertex operator:
\begin{equation}
\delta V = \oint \lambda (z') T (z') V(z)
\end{equation}
The world-sheet energy momentum tensor includes two parts,
\begin{eqnarray}
& T (z') =  T^m (z')+  T^g (z') \nonumber \\ & T^m (z) =
-\frac{1}{4\alpha'} \partial X_{\mu} \partial X^{\mu} \nonumber
\\ & T^g (z) = (\partial b ) c - 2\partial (b c)
\end{eqnarray}
Using the operator products of ghost $c$ and antighost $b$,
\begin{equation}
b(z_1) c(z_2) \sim \frac{1}{z_{12}}
\end{equation}
and the equation of motion (\ref{motion equ}),
\begin{equation}
\delta V = \lambda \partial V + \alpha' k^2 (\partial \lambda) V
\end{equation}
So the finite transformation of the operator is
\begin{equation}
V' (z') = {\Big ( }\frac{d z'}{d z}{\Big )}^{-\alpha' k^2} V (z)
\end{equation}
In string field theory, an arbitrary 3-point vertex can be defined by
\begin{equation}
\langle V[ h_1 (0) ]  V[ h_2 (0) ] V [h_3 (0) ] \rangle = \langle
h_1 [ V(0) ] h_2 [ V(0) ] h_3 [ V(0) ] \rangle
\end{equation}
with $h_i$ arbitrary maps of $z=0$ into the upper complex
plane~\cite{E. peskin:1988}. This differs from the previous expression only by terms with extra factors of $k_i^2$.  By the previous argument (\ref{ksquared}), such terms vanish by the gauge-invariant equation of motion.

\section{Neveu-Schwarz vertex}\label{section4}

This formalism can also be generalized to the NSR string: Start
from the integrated vertex operator $\oint \epsilon \cdot
D_{\theta} X e^{ik \cdot X (Z) }$ with $Z = (z, \theta)$,
$X^{\mu}(Z) = x^{\mu} (z) + i \theta \psi (z)$ and $D \equiv
D_{\theta} = \partial_{\theta} + \theta \partial_z$. Here we will
use the language for the Neveu-Schwarz string of the ``Big
Picture"~\cite{bigpicture}. Then as in the bosonic case, if the
commutator can be written as $[Q, W \} = D_{\theta} V$, then $[Q,
V \} =0$. To simplify the calculation, we replace $X$ by
$(\alpha'/2)^{1/2} X$. Then
\begin{equation}
W = (\alpha'/2)^{1/2} \epsilon \cdot D_{\theta} X e^{ik \cdot
(\alpha'/2)^{1/2} X (Z) }
\end{equation}
 and the propagator
\begin{equation}\label{propagator}
 X^{\mu}(z',
\theta') X^{\nu}(z, \theta) \sim -4 ln |z'- z - \theta' \theta |
\eta^{\mu \nu}
\end{equation}
We also write $\sqrt{\alpha'/2}\thinspace \epsilon$
as $\epsilon$ and $\sqrt{\alpha'/2}\thinspace  k$ as $k$ and restore these factors in the final result.

The BRST operator $Q$ in the NSR string is
\begin{equation}
Q = \frac{1}{2\pi i} \oint d z' d \theta' (C T^x + \frac{1}{2} C
T^g)
\end{equation}
with
\begin{equation}\label{Tx}
T^x = - \frac{1}{8} C(z', \theta') (D_{\theta'} X^{\mu})
(\partial' X_{\mu})
\end{equation}
and
\begin{equation}\label{Tg}
T^g = \frac{1}{2} (D_{\theta'} B) (D_{\theta'} C ) - \frac{3}{2} B
(\partial' C) - (\partial' B) C
\end{equation}
Here $C = c + \theta \gamma$ and $B = \beta + \theta b$, where $c$ and
$b$ are anticommuting superconformal ghost and antighost, and $\beta$
and $\gamma$ are commuting superconformal ghost and antighost.

Because $W$ contains no ghosts, it is only necessary to
compute
\begin{equation}
[Q, W \} =  \frac{1}{2\pi i} \oint d z' d \theta' (- \frac{1}{8})
[ C(z', \theta') (D_{\theta'} X^{\mu}) (\partial' X_{\mu})]
[\epsilon \cdot D_{\theta} X e^{ik \cdot X (Z) } ]
\end{equation}
Using
\begin{equation}
\oint d z' d \theta' \frac{1}{z'-z-\theta' \theta} [f(z') +
\theta' g(z') ] = D_{\theta} [f(z) + \theta g(z)]
\end{equation}
and
\begin{equation}
\oint d z' d \theta' \frac{\theta' - \theta}{z' -z} [f(z') +
\theta' g(z') ] = f(z) + \theta g(z)
\end{equation}
the result for $V$ is
\begin{eqnarray}
V & = & - D_{\theta} [C (\epsilon \cdot D_{\theta} X) e^{ik \cdot
X (Z) } ] + \frac{1}{2} (D_{\theta} C) (D_{\theta} X \cdot
\epsilon) e^{ik \cdot X (Z) } \nonumber \\ & & - 2 i
(\epsilon \cdot k) (\partial C) e^{ik \cdot X (Z)} \nonumber \\ &
= & G + H
\end{eqnarray}
where $G$ represents the first two terms and $H$ represents the last
term.

To see the gauge invariance of the amplitude, we should know the
gauge transformation of the vertex operator $\delta V$. Writing
the vertex operator as:
\begin{equation}
V =  - D_{\theta} [C A^{\mu} (D_{\theta} X_{\mu})] - 2 (\partial
C) (\partial_{\mu} A^{\mu}) + \frac{1}{2} (D_{\theta} C) A^{\mu}
(D_{\theta} X_{\mu}) ,
\end{equation}
then the gauge transformation of the vertex operator is
\begin{eqnarray}
\delta V & = & - D_{\theta} [C (\partial^{\mu} \lambda) (
D_{\theta} X_{\mu})] - 2 (\partial C) [\partial_{\mu}
(\partial^{\mu} \lambda)] + \frac{1}{2} (D_{\theta} C) (D_{\theta}
X_{\mu}) (\partial^{\mu} \lambda) \nonumber \\ & = & - \frac{1}{2}
(D_{\theta} C) ( D_{\theta} X_{\mu} ) \partial^{\mu} \lambda - 2
(\partial C) \bo \lambda + C (\partial X_{\mu})
\partial^{\mu} \lambda \nonumber \\ & & - C (D_{\theta} X_{\mu}) (D_{\theta} X_{\nu})
\partial^{\nu} \partial^{\mu} \lambda
\end{eqnarray}
under $A \rightarrow A + \partial_{\mu} \lambda$. The last term is
zero because $D_{\theta} X_{\mu}$ and $
D_{\theta} X_{\nu}$ anticommute. $\delta V$ can be written as the commutator of the BRST
operator and the function $\lambda$:
\begin{eqnarray}
[ Q, \lambda (X(z,\theta)) \} & = & \frac{1}{2\pi i} \oint d z' d
\theta' \frac{1}{8} C(z', \theta') (D_{\theta'} X^{\mu})
(\partial' X_{\mu}) \lambda (X(z,\theta) \nonumber \\ & = &
-\frac{1}{2} D_{\theta} (C  D_{\theta} X_{\mu} ) \partial^{\mu}
\lambda - 2 (\partial C) \bo \lambda + \frac{1}{2} C (\partial
X_{\mu}) \partial^{\mu} \lambda
\end{eqnarray}
So the amplitude $\langle V V V \oint W \cdots \oint W \rangle$ is
gauge invariant just as in the bosonic case.

\section{Neveu-Schwarz three-point amplitude}

Similarly to the bosonic case, we are going to use this vertex operator in the NSR
string to compute the 3-point amplitude between 3 YM gauge
bosons. First of all, we should know the correlation function
\begin{eqnarray}
& & \langle 0 | C(z_1, \theta_1) C(z_2, \theta_2) C(z_3, \theta_3)
| 0 \rangle \nonumber \\ & = & \theta_1 \theta_2 z_3 (z_1 + z_2) +
\theta_2 \theta_3 z_1 (z_2 + z_3) + \theta_3 \theta_1 z_2 (z_3 +
z_1) \nonumber
\end{eqnarray}
Then using the propagator between $X$'s as (\ref{propagator}), the
anticommutation relation
between $C$'s and $D_{\theta}$'s, and the gauge-covariant equation of motion (\ref{motion
equ}), after a more-complicated calculation, to the lowest order in
$\alpha'$, we find
\begin{eqnarray}
 {\cal A}_3 & = & \frac{2g_{YM}}{\alpha'^2} \langle
V (z_1, \theta_1)
V (z_2, \theta_2)
V(z_3, \theta_3) \rangle \nonumber
\\ & = & i g_{YM} (2\pi)^D \delta^D (\Sigma_i k_i) [
(\epsilon_1 \cdot \epsilon_2) (\epsilon_3 \cdot k_{12}) +
(\epsilon_2 \cdot \epsilon_3) (\epsilon_1 \cdot k_{23}) +
(\epsilon_3 \cdot \epsilon_1) (\epsilon_2 \cdot k_{31}) ] \nonumber \\
\end{eqnarray}
Again this
amplitude corresponds to the $Tr(F^{\mu \nu} F_{\mu \nu})$ term in
the YM theory.

This amplitude is independent of the
anticommuting coordinate $\theta$, as expected. It is also
independent of $z$, as in the bosonic case. We will discuss this more
later. One thing different from the bosonic case is that $F^3$ terms cannot be supersymmetrized. This corresponds
to the absence of $k^3$ terms in the NSR string amplitude. To
verify this, it is necessary to compute to the second order in
$\alpha'$ in the amplitude.

To second order in $\alpha'$, factors like
\begin{equation}\label{expansion}
|y_{ij}-\theta_i \theta_j|^{2\alpha' k_i \cdot k_j}, \quad
D_{\theta_k} |y_{ij}-\theta_i \theta_j|^{2\alpha' k_i \cdot k_j},
D_{\theta_k} D_{\theta_l} |y_{ij}-\theta_i \theta_j|^{2\alpha' k_i \cdot
k_j}, \cdots
\end{equation}
are involved in the amplitude  ${\cal A}_3$. To lowest order in $\alpha'$,
the first factor is just one, the rest zero.
But the expansion of $2\alpha' k_i \cdot k_j$ contributes to the second order in $\alpha'$, as do
\begin{eqnarray}
& & \langle G (y_1) G (y_2) G (y_3) \rangle , \nonumber
\\ & & \langle  G (y_1) G (y_2) H (y_3) \rangle , \cdots ,
\nonumber \\ & & \langle G (y_1) H (y_2) H (y_3) \rangle , \cdots
\end{eqnarray}
The term from
\begin{eqnarray}
\langle H(y_1) H(y_2) H(y_3) \rangle
\end{eqnarray}
vanishes.

We calculate the expansion of $|y_{ij}-\theta_i
\theta_j|^{2\alpha' k_i \cdot k_j}$ and its derivatives $D_{\theta_k}
|y_{ij}-\theta_i \theta_j|^{2\alpha' k_i \cdot k_j}$, etc., using
\begin{eqnarray}
& & |y_{12}-\theta_1 \theta_2|^a |y_{23}-\theta_2 \theta_3|^b
|y_{31}-\theta_3 \theta_1|^c \nonumber \\ & = & [ 1 -
\frac{a}{y_{12}} \theta_1 \theta_2 - \frac{b}{y_{23}} \theta_2
\theta_3 - \frac{c}{y_{31}} \theta_3 \theta_1 ] |y_{12}|^a
|y_{23}|^b |y_{31}|^c \nonumber \\ & = & [ D_{\theta_1}
(|y_{12}|^a \theta_{12}) ] [ D_{\theta_2} (|y_{23}|^b
\theta_{23})] [ D_{\theta_3} (|y_{31}|^c \theta_{31})]
\end{eqnarray}
to the first order in $a, b, c$.
Then
\begin{eqnarray}
& & D_{\theta_1} (|y_{12}-\theta_1 \theta_2|^a |y_{23}-\theta_2
\theta_3|^b |y_{31}-\theta_3 \theta_1|^c ) \nonumber \\ & = &
(\partial_1 |y_{12}|^a) |y_{23}|^b |y_{31}|^c \theta_{12} +
|y_{12}|^a |y_{23}|^b (\partial_3 |y_{31}|^c) \theta_{31}
\nonumber \\ & & + \theta_1 \theta_2 \theta_3 [ (\partial_1
|y_{12}|^a) \partial_3 (|y_{23}|^b |y_{31}|^c) + \partial_2
(|y_{12}|^a |y_{23}|^b) (\partial_3 |y_{31}|^c) ]
\end{eqnarray}
\begin{equation}
D_{\theta_1} D_{\theta_2} ( |y_{12}-\theta_1 \theta_2|^a
|y_{23}-\theta_2 \theta_3|^b |y_{31}-\theta_3 \theta_1|^c ) =
\frac{a}{y_{12}} + \frac{a}{y_{12}^2} \theta_1 \theta_2 , \cdots
\end{equation}
The detailed calculation is too long to show here. But combining
both contributions, the 3-point amplitude to the second order in
$\alpha'$ is just zero as predicted.

To study the vertex position dependence of the amplitude, we compute the
conformal transformation of $V$
\begin{equation}
\delta V  = \frac{1}{2\pi i} \oint d z' d \theta' \lambda (T^x +
T^g) (z', \theta') V (z,\theta)
\end{equation}
with $T^x$ and $T^g$ defined in eqs. (\ref{Tx}-\ref{Tg}).
Using the OPE of $B$ and $C$
\begin{equation}
B(z_1, \theta_1) C(z_2, \theta_2) \sim \frac{\theta_1 -
\theta_2}{z_1-z_2} ,
\end{equation}
\begin{equation}
\delta V = (2 k^2) (\partial \lambda) V + \frac{1}{2} (D_{\theta})
(D_{\theta} V) + \lambda (\partial V)
\end{equation}
Replacing $k$ by $\sqrt{\alpha'/2}\thinspace k$ as mentioned in the
beginning of section \ref{section4}, and taking the infinitesimal
superconformal transformation $\lambda = 2 \theta \eta$ with
$\eta(z)$ anticommuting,
\begin{equation}
\delta V = (2 \alpha' k^2) (\theta \partial \eta) V + \eta
Q_{\theta} V
\end{equation}
where $Q_{\theta} = \partial_{\theta} - \theta \partial_z$.

Thus the vertex operator has the weight $\alpha' k^2$ and
transforms as
\begin{equation}
V'(z', \theta') = (D_{\theta} \theta')^{-2\alpha' k^2} V(z,\theta)
\end{equation}
So the amplitude transforms as
\begin{eqnarray}
& & \langle V'(z'_1,\theta'_1) V'(z'_2,\theta'_2) V'(z'_3,\theta'_3)
\rangle \nonumber \\ & = & (D_{\theta_1} \theta'_1)^{-2\alpha'
k_1^2} (D_{\theta_2} \theta'_2)^{-2\alpha' k_2^2} (D_{\theta_3}
\theta'_3)^{-2\alpha' k_3^2} \langle V(z_1,\theta_1)
V(z_2,\theta_2) V(z_3,\theta_3) \rangle
\end{eqnarray}
The expansion of $\alpha' k_i^2$ gives just one, using $\bo F =0$.
This implies
\begin{eqnarray}
\langle V'(z'_1,\theta'_1) V'(z'_2,\theta'_2) V'(z'_3,\theta'_3)
\rangle = \langle V(z_1,\theta_1) V(z_2,\theta_2) V(z_3,\theta_3)
\rangle
\end{eqnarray}
and no higher order terms in $\alpha'$.

\section{Conclusions}

We have given a general construction for gauge-covariant vertex operators, and applied it to the YM vertex in the string and superstring in 3-point amplitudes.  This method allows direct calculation of gauge-invariant results, analogous to nonlinear sigma models, and can also be applied to string field theory.  Possible future applications include higher-point amplitudes.

\section{Acknowledgement}

This work is supported in part by National Science Foundation
Grant No.\ PHY-0098527.

\end{document}